\documentclass[aps,12pt,nofootinbib,preprint]{revtex4}
\pdfoutput=1
\usepackage{textcomp}
\usepackage{amssymb}
\usepackage{anysize}
\usepackage{bold-extra}
\usepackage{enumerate}
\usepackage{graphicx}
\usepackage{lscape}
\usepackage{mathrsfs}
\usepackage{multirow}
\usepackage[final]{pdfpages}
\usepackage{rotating}
\usepackage{subfig}
\usepackage{url}
\usepackage{wrapfig}
\usepackage{amsmath}
\usepackage{color}
\usepackage{fancyhdr}
\usepackage{verbatim} 
\usepackage[linktocpage,colorlinks]{hyperref}
\def\nn{\nonumber}
\def\bea{\begin{eqnarray}}
\def\eea{\end{eqnarray}}
\def\ba{\begin{eqnarray}}
\def\ea{\end{eqnarray}}
\def\be{\begin{equation}}
\def\ee{\end{equation}}
\def\beq{\begin{equation}}
\def\eeq{\end{equation}}

\def\lsim{\mbox{\raisebox{-.6ex}{~$\stackrel{<}{\sim}$~}}}

\def\({\left(}
\def\){\right)}
\def\vev#1{\langle #1\rangle}

\def\OMIT#1{{\color{red}{#1}}}
\begin{document}

\preprint{UCSD/PTH 13-06}
\title{Searching for New Physics in the Three-Body Decays of the Higgs-like Particle}
\author{Benjam\'{i}n Grinstein}
\email{bgrinstein@ucsd.edu}
\affiliation{Department of Physics, University of California, San Diego, La Jolla, CA 92093 USA}
\author{Christopher W. Murphy}
\email{cmurphy@physics.ucsd.edu}
\affiliation{Department of Physics, University of California, San Diego, La Jolla, CA 92093 USA}
\author{David Pirtskhalava}
\email{pirtskhalava@physics.ucsd.edu}
\affiliation{Department of Physics, University of California, San Diego, La Jolla, CA 92093 USA}

\begin{abstract}
  We show that the three-body decays of the resonance recently
  discovered at the LHC are potentially sensitive to effects of new
  physics.  Even if the fully integrated partial decay widths are
  consistent with the minimal Standard Model there is information that
  is lost upon integration, which can be uncovered in the differential
  decay widths.  Concentrating on the decay $h \to Z \ell \bar{\ell}$,
  we identify the regions in the three-body phase space in which these
  effects become especially pronounced and could be detected in future
  experiments.
\end{abstract}

\maketitle

\section{Introduction}
The first and simplest realization of the electroweak symmetry
breaking sector of the Standard Model (SM) consists of a single,
self-interacting, electroweak-doublet complex scalar field. The
longitudinal components of the $W^\pm$ and $Z$ gauge bosons are in
fact three of the four real components of the complex scalar
doublet. The remaining physical excitation of the doublet is the
``Higgs particle.'' There is compelling evidence for the assumption
that the Higgs particle is the 125 GeV resonance observed at LHC
experiments~\cite{Aad:2012tfa,Chatrchyan:2012ufa}. Not only are the
production rates and partial widths of the resonance very close to what
this model predicts~\cite{Espinosa:2012ir,Carmi:2012in,Ellis:2013lra,Giardino:2013bma}, but there is no evidence for other resonances
despite extensive searches.

Other, more involved models can equally well or better
reproduce the production and decay widths of the new resonance. They come with additional states that may have somehow escaped
detection. For example, the simplest extensions of the one doublet
model are  models with two Higgs doublets.  These models have been
shown to reproduce the LHC 125 GeV-resonance data~\cite{Altmannshofer:2012ar,Bai:2012ex,Drozd:2012vf,Chang:2012zf,Celis:2013rcs,Chiang:2013ixa,Barroso:2013awa},  analyses have
been presented with the rates of production and decay of the
additional resonances, and strategies have been delineated to 
search for them~\cite{Grinstein:2013npa}. However, direct detection of new states may be
impossible given the limited energy and performance of LHC
experiments.

\begin{figure}
  \centering
  \includegraphics[width=0.5\textwidth]{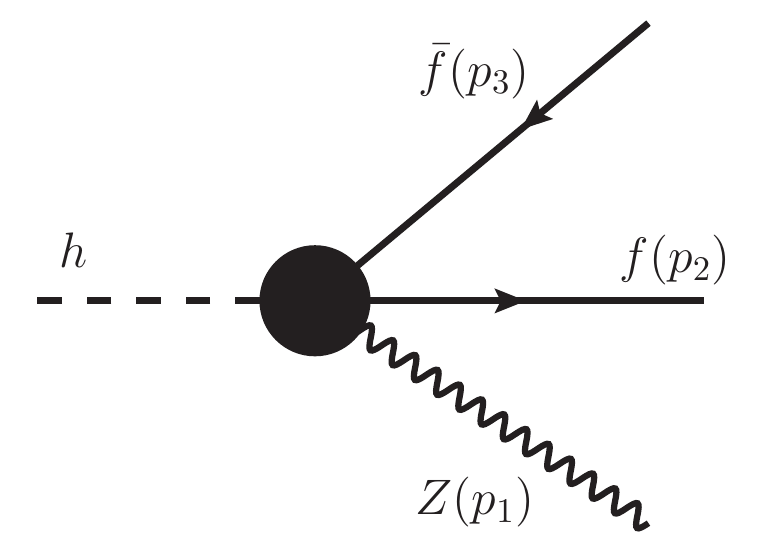} \\
  \caption{Three-body decay $h\to Z f\bar f$; the main focus of the
    present work is for the case that the final fermion pair is
    comprised of charged leptons, $f\bar f = \ell\bar\ell$.}
 \label{fig:diagram6finalstate}
  % \end{figure}\begin{figure}
  %\centering
  %\includegraphics[width=0.5\textwidth]{Untitled.pdf} \\
   %\caption{Feynman diagram for the process being studied.}
 %\label{fig:diagram4finalstate}
  % \end{figure}\begin{figure}
  %\centering
  %\includegraphics[width=0.5\textwidth]{Untitled.pdf} \\
   %\caption{Feynman diagram for the process being studied.}
 %\label{fig:diagram}
\end{figure}

In the absence of direct evidence for new states one may resort to
precision measurements of the properties of the 125 GeV resonance to
falsify or strengthen the minimal model assumption. Here we have in
mind a variety of colliders that have been
proposed~\cite{Alsharoa:2002wu,Djouadi:2007ik,Linssen:2012hp} that
could serve as ``Higgs factories,'' which we presume allow for
precision measurements of the type we discuss below. Experience with
hadronic physics would suggest that the nature of the 125 GeV
resonance can be explored by measuring form factors.  Here instead we
settle for a less ambitious program, namely to study the effects of
new physics that can be characterized by an effective Hamiltonian
through the spectrum in its three-body decays, see
Fig.~\ref{fig:diagram6finalstate}. We assume that the production
rates and the decay branching fractions are well measured, or will be
well measured by the end of the 2015 run of the LHC experiments, and
furthermore that they are in good agreement with the predictions of
the minimal SM. We will show that there is information that is lost in
the fully integrated partial decay widths that can be uncovered in the
differential decay widths.

Recently, the use of differential distributions to confirm or reject
various hypotheses about the underlying physical processes has been
successful.  Specifically, the differential decay rate for $h \to
4\ell$~\cite{Chang:1993jy,Skjold:1994qn,Arens:1994wd,Grzadkowski:1995rx,Buszello:2002uu,Choi:2002jk,Djouadi:2005gi,Bredenstein:2006rh,Bredenstein:2006ha,Godbole:2007cn,Cao:2009ah,Christensen:2010pf,DeRujula:2010ys,Avery:2012um,Choi:2012yg,Bolognesi:2012mm} has been
used to determine the spin and parity~\cite{Chatrchyan:2012jja,CMS-PAS-HIG-13-002,
  ATLAS-CONF-2013-013} of the recently discovered resonance at the
LHC, zero and even respectively.\footnote{The tensor structure of this new scalar's couplings gauge bosons can be determined in the same manner~\cite{Stolarski:2012ps}.}  The error bars on the differential
measurements are currently large, of order 100\% of the central
values, which is large enough to hide new physics that doesn't modify
the total rate significantly, but whose shapes may differ wildly from
the SM prediction. We will show in Sec.~\ref{sec:dis} that this
can occur in realistic extensions of the SM.

\section{Generalities}
Ignoring the effects of binding and confinement it is easy to classify
and list the possible channels for three-body decays of the Higgs. The
three particle state has either three gauge bosons or one gauge boson
and two fermions, either two quarks or two leptons. In this work we
study decays into two fermions and a gauge boson. We further restrict
the state to the case that the two fermions are charged leptons since
these offer the best prospect for high precision measurements. Then
the accompanying gauge boson is neutral and we concentrate primarily
on the case that it is a $Z$-boson, that is, on decays of the form
$h\to Z\ell\bar\ell$.  The decay $h\to\gamma\ell\bar\ell$ arises in
the SM from $h\to\gamma Z^*\to \gamma \ell\bar\ell$ and $h\to
\gamma\gamma^*\to\gamma \ell\bar\ell$ at one loop. Because it is
suppressed in the SM it may be interesting to study this process, since
new physics may produce large deviations from the SM
prediction. However, $h\to\gamma Z^*\to \gamma \ell\bar\ell$ is
primarily a 2-body decay of $h$ and does not have the non-trivial kinematic
information that is the focus of our investigation, while $h\to
\gamma\gamma^*\to\gamma \ell\bar\ell$ is constrained by the
$h\to\gamma\gamma$ branching fraction.

Consider supplementing the Standard Model with the following operators
which can contribute to $h \rightarrow Z\, \ell\, \bar{\ell}$ at tree
level
\begin{align} 
\label{eq:lag}
\mathcal{L}_{NP} &= \frac{e}{s_W c_W}\left( \frac{c_{\ell Z}}{4 \pi v} \bar{\ell} \sigma^{\mu\nu} \ell Z_{\mu\nu} + \bar{\ell}\gamma^{\mu}(c_L P_L + c_R P_R)\ell Z_{\mu}\right)\frac{h}{v} \\
&+ \frac{\alpha}{4 \pi}\left(\frac{c_{ZZ}}{s_W^2 c_W^2} Z_{\mu\nu}
  Z^{\mu\nu} + \frac{c_{Z\gamma}}{s_W c_W} Z_{\mu\nu}
  F^{\mu\nu}\right)\frac{h}{v}. \nn
\end{align}
We will refer to the second operator of the first line as ``the
current interaction,'' labeling it by $O_{ZJ}$.  The peculiar
normalization of the various terms was chosen such that all coefficients, $c$, are of order one according to naive dimensional analysis~\cite{Manohar:1983md,Jenkins:2013sda} when the cutoff of the effect theory is taken to be $\Lambda = 4 \pi v$. The Lagrangian \eqref{eq:lag} is a
parametrization of possible interactions beyond those in the SM,
regardless of the details of underlying new dynamics about which we
remain agnostic. The dimension-3 operator $hZ^\mu Z_\mu$ has not been
included. This operator would renormalize the already existing
tree-level SM coupling. We will briefly discuss the effects of such a
renormalization below. The dipole terms in \eqref{eq:lag} can not
arise from the flavor $SU(3)^5$-invariant UV theory that we assume in
the present work. Nevertheless, we include their contributions for
completeness.\footnote{In any case, the effects of these operators for
  the processes we are considering should be expected to be rather
  weak since they do not interfere with the leading SM diagram.}

While we do not assume a particular UV-completion of the low energy
theory, let us discuss how the interactions in \eqref{eq:lag} may be
properly embedded in a model in which the $Z/W$ bosons acquire mass
through spontaneous breaking of electroweak symmetry. This is
necessary because while only Higgs data provide constraints on the
coefficients of the operators in Eq.~\eqref{eq:lag}, additional
indirect constraints arise when these operators are related to others
by electroweak symmetry.

There are two approaches to an effective Lagrangian for beyond the SM
interactions of the Higgs particle, with the electroweak gauge
symmetry linearly or non-linearly realized. In the linear realization,
the above mentioned effective Lagrangian consists of a sum of
operators of dimension higher than four, including the scalar doublet
that can be constructed in terms of the fields in the SM. This can be
done systematically by assuming diminishing effects of operators of
increasing dimensions. We will therefore concentrate on operators of
dimension no bigger than six. In the linear realization such operators
were first classified in Ref.~\cite{Buchmuller:1985jz}. The minimal
set, which excludes a few redundant operators, is presented in
Ref.~\cite{Grzadkowski:2010es}. The strength of each operator is
characterized by its coefficient in the effective Lagrangian. Some of
these coefficients are severely constrained both by flavor physics,
e.g., neutral meson mixing, and by electroweak precision data
(EWPD). We assume the new physics is CP conserving and flavor
symmetric, that is, the operators in $\mathcal{L}_{NP}$ are invariant
under $SU(3)^5$ transformations of the quark and lepton doublet and
singlet fields of the SM. Hence the constraints from flavor physics
are immediately alleviated.  The effective theory analysis of
precision electroweak data has a long history
\cite{Buchmuller:1985jz,Grinstein:1991cd}, one particular example
being the so-called oblique corrections
\cite{Peskin:1990zt,Holdom:1990tc,Altarelli:1990zd,Golden:1990ig,Peskin:1991sw}
(see \cite{Maksymyk:1993zm,Barbieri:2004qk} for extensions of the
latter formalism). More generally, the constraints from electroweak
precision measurements on the resulting effective Lagrangian have been
studied in \cite{Barbieri:1999tm, Han:2004az}.  Of particular interest
for our work is the analysis by Han and Skiba \cite{Han:2004az}
(henceforth HS). They consider a basis of 21 dimension-six operators
and determine the simultaneous constraints on the corresponding
21-dimensional space of coefficients of these operators. Of these 21
operators, 11 are four-fermion operators and play no role in Higgs
decays (in low orders in perturbations theory). Of the remaining
operators, 9 are bi-linear in the scalar-doublet field $H$ and one is
cubic in the field strength tensor of the weak-SU(2) gauge field. HS
give an expression for $\chi^2$ as a function of the 21 parameters,
so one can find the constraints on the coefficients even after
assuming that some vanish, say, by some theoretical argument (or
prejudice). The simultaneous fit to all 21 parameters gives very loose
bounds on any one of the parameters (which however are highly
correlated). For example, the familiar oblique parameters are found to
be $S=53\pm250$ and $T=36\pm223$.\footnote{The HS fit assumes
  $M_h=113$~GeV. We have used their fit without the modifications
  required to account for the observed value of $M_h\sim125$~GeV.  The
  correction is insignificant. For example, the change in the central
  value of $S$ is $\delta S= 0.005$.}  By contrast, the same fit
assuming all but $S$ and $T$ vanish yields $S=0.05\pm0.10$ and
$T=-0.04\pm0.10$.  More generally, the same situation occurs for any
fit only including a particular subset of the parameter
space. Assuming vanishing of the coefficients of the operators that
complete it to the full set, implies constraints on the given subset
that are more severe than they would be for the simultaneous fit
including all of the independent operators.

The very loose HS
bounds are in fact in conflict with recent LHC bounds on contact
interactions from lepton pair production
\cite{Chatrchyan:2012hda,Aad:2012bsa}. Being conservative we set the
coefficient of all four-fermion operators in the HS basis to zero. The
remaining operators are 
\begin{gather*}
O_{WB}=H^\dagger \sigma^a H W^a_{\mu\nu}B^{\mu\nu}, \quad
O_h=|H^\dagger D_\mu H|^2,\\ O^s_{hl}=H^\dagger iD_\mu
H\bar l\gamma^\mu l+\text{h.c.},\quad
O^t_{hl}=H^\dagger\sigma^a iD_\mu
H\bar l\sigma^a\gamma^\mu l+\text{h.c.},
\\ O^s_{hq}=H^\dagger iD_\mu H\bar q\gamma^\mu q+\text{h.c.},\quad
O^t_{hq}=H^\dagger\sigma^a iD_\mu H\bar q\sigma^a\gamma^\mu
q+\text{h.c.},\\
O_{hu}=H^\dagger iD_\mu H\bar u\gamma^\mu u+\text{h.c.},\quad
O_{hd}=H^\dagger iD_\mu H\bar d\gamma^\mu d+\text{h.c.},\\
O_{he}=H^\dagger iD_\mu H\bar e\gamma^\mu e+\text{h.c.}\quad\text{and}\quad
O_W=\epsilon^{abc}{W^a_\mu}^\nu{W^b_\nu}^\lambda{W^b_\lambda}^\mu.
\end{gather*}
Here $l$ and $q$ are left-handed electroweak doublets while $u$, $d$
and $e$ are right handed electroweak singlets.  The resulting best fit
values of the coefficients $a_i$ of these operators in the Lagrangian,
$\Delta\mathcal{L}_{\text{eff}}=\sum_i a_i O_i$, are given in
Table~\ref{HSas}. The HS basis excludes operators that while being CP
invariant and flavor symmetric, do not contribute to the processes
considered in electroweak precision tests. 

The operators that may contribute to $h\to Z\ell\bar\ell$, according
to the way in which they are constrained, can roughly be divided into
two types: a) the ones that are mostly constrained by the current
Higgs data (such as the $h\to \gamma\gamma$ rate), and b) those that
are mostly constrained by EWPD. Examples of dimension-6 operators from
category a), are
\[
O_{BB}=H^\dagger H B_{\mu\nu} B^{\mu\nu}, \quad O_{WW}= H^\dagger H ~\text{Tr}[W^{\mu\nu}W_{\mu\nu}]~.  
\] 
The coefficients of these  operators are not constrained by electroweak data, since their sole contribution to EW precision observables is to rescale the gauge kinetic terms. However, one combination of these does
contribute to the $h\to\gamma\gamma$ rate and is therefore tightly
constrained by LHC measurements. Operators from the category b) are those in the HS basis. 
Finally, there are dimension-6 operators involving only quark and/or gluon fields, that are constrained neither by the current Higgs data, nor
by electroweak precision experiments. The latter are irrelevant for the present work.

\begin{table}[t]
\begin{tabular}{|c|c|c|c|c|c|c|c|c|c|}
\hline
$a_{WB}$&$a_{h}$&$a^s_{hl}$&$a^t_{hl}$&$a^s_{hq}$&$a^t_{hq}$&$a_{hu}$&$a_{hd}$&$a_{he}$&$a_{W}$\\
\hline
$4.6\pm7.5$ &$ 0.0\pm 26.$ &$ 2.8\pm6.7$&$ 0.9\pm21.$ &$-0.9\pm2.2$
&$0.9\pm21.$&$-3.6\pm8.9$&$1.7\pm4.4$&$5.6\pm13.$&$-3.9\pm 32.$\\
\hline
\end{tabular}
\caption{Best fit values and 1$\sigma$ errors, in units of TeV$^{-2}$,
  of the coefficients of dimension 6 operators in the HS basis when
  the coefficient of four-fermion operators are assumed to vanish.}
\label{HSas}
\end{table}

Expanding the operators from the HS basis about the electroweak
symmetry breaking vacuum, one obtains various contributions to the
couplings shown in \eqref{eq:lag}. For example, the coefficients
$c_L$, $c_R$
and $c_{Z\gamma}$ are given in terms of the coefficients of the dimension-6
operators from the HS basis as follows 
\begin{equation}
\label{eq:tbase}
c_L = - v^2 (a_{h\ell}^s + a_{h\ell}^t),\quad c_R = - v^2 a_{h e},\quad \frac{\alpha c_{Z\gamma}}{4 \pi s_W c_W} = - v^2 (c_W^2 - s_W^2) a_{WB}~.
\end{equation}
As already emphasized, a simultaneous fit of all relevant
flavor-preserving operators to the electroweak data yields relatively
weak constraints on the Higgs couplings in \eqref{eq:lag}.

A question naturally arises as to whether it is possible, let alone
reasonable, to have UV completions of the SM that result in large
coefficients of the effective interactions in \eqref{eq:lag} without
automatically (without fine tuning) producing conflicts with
EWPD. 
%\color{red}{Supplementing the SM with a heavy $W^{\prime}$ with SM
%couplings, generates coefficients $c_{\gamma\gamma} = -7$ and
%$c_{Z\gamma} = -3.55$. It may worth mentioned that somewhere around   here.}\color{black}\, 
Manohar has shown, in a simple extension of the SM with a
self-interacting set of SO(N)-symmetric scalar SU(2)-doublets, that
this is indeed the case \cite{Manohar:2013rga}; see also
\cite{Jenkins:2013fya}. While we do not review Manohar's model here,
we do summarize his findings: (i) A renormalizable model extension of
the SM produces only four dimension-6 operators in its low-energy EFT,
namely $O_{WB}$, $O_{WW}$, $O_{BB}$, and $O_{W}$; (ii) The
coefficients of these operators in the EFT are independent, except for
the relation $a_{BB}=\tan^2(\theta_w)Y^2a_{WW}$, with $Y$ a free
parameter; (iii) At strong coupling the coefficients in the EFT are
not suppressed by the loop factor $1/16\pi^2$.  The coefficients
$a_{WW}$ and $a_{BB}$ are not constrained by electroweak precision
data (EWPD). In Manohar's model there are only two entries in the
Han-Skiba EWPD fit, and for them we obtain $a_{WB}=(-0.1\pm2.0)\times
10^{-2}~\text{TeV}^{-2}$ (corresponding to Peskin-Takeuchi's $S=0.0\pm
0.3$) and $a_{W}=(-0.2\pm3.8)~\text{TeV}^{-2}$. The unconstrained
coefficients $a_{WW}$ and $a_{BB}$ can modify the rates for
$h\to\gamma\gamma$ and $h \to Z\gamma$ independently and with
significant enhancements in one or the other or both. Hence consistent
models of NP can produce the effects we explore here without recourse
to fine-tuning.

A situation in which EWPD constraints can be even weaker occurs when
the UV theory realizes the electroweak symmetry nonlinearly.
The non-linear realization is appropriate for models in which the symmetry is spontaneously broken dynamically, which can occur for example by a
new hyper-strong interaction with characteristic scale
$\Lambda\lesssim 4\pi v$~\cite{Appelquist:1993ka}. The Higgs particle
in this setup is an accidentally light ($m_h\ll \Lambda$) CP-even
scalar. That the properties of this scalar resemble those of the SM
Higgs particle in the first approximation arises naturally if the scalar
is in fact a pseudo-dilaton of the nearly conformal hyper-strong
interaction \cite{Goldberger:2007zk}. In this case, the effective
Lagrangian is constructed from SM fields, as well as a field $h$
for the Higgs-like particle, and a triplet of would-be-Goldstone bosons
$\pi^a$ from electroweak symmetry breaking that are eaten by the $W$
and $Z$ fields via the Higgs mechanism. While the electroweak
transformation of the triplet is non-linear, the effective
Lagrangian is easily constructed \cite{Grinstein:2007iv} using the
Callan-Coleman-Wess-Zumino prescription \cite{Callan:1969sn} for a
field $\Sigma=\exp{i\sigma^a\pi^a/v}$ that transforms linearly.

One can readily see how the effective Lagrangian~\eqref{eq:lag} can arise from a
UV theory with a non-linear realization of $SU(2)_W\times U(1)_Y$. Gauge invariant
operators can be constructed from objects such as the singlet Higgs field $h$, the nonlinear sigma model field $\Sigma$, its covariant
derivative, $D_\mu\Sigma=\partial_\mu\Sigma-i\frac{g}{2} W^a_\mu\sigma^a \Sigma-i \frac{g'}{2}\Sigma\sigma^3 B_\mu$, gauge field strengths $W^a_{\mu\nu}$, $B_{\mu\nu}$, etc. Under the electroweak group, the $\Sigma$ field transforms as $\Sigma \to U_L\Sigma U^{\dagger}_Y$, where $U_L$ is the $SU(2)_{W}$ transformation matrix, while $U_Y=\text{exp}(i\alpha_Y\sigma^3/2)$ corresponds to the hypercharge transformation. Taking into account the transformation property of the $SU(2)$ field strength, $W_{\mu\nu}\to U_L W_{\mu\nu}U_L^\dagger$, one can easily construct gauge invariant operators that in the unitary gauge ($\Sigma=1$) reproduce the terms in the Lagrangian \eqref{eq:lag}. For example, the operators $O_{Z\gamma}$ and $O_{ZZ}$ can be obtained from the following gauge invariant operators
\be
\label{eq:nl1}
h\text{Tr} [W_{\mu\nu} W^{\mu\nu}], \quad h B_{\mu\nu}B^{\mu\nu}, \quad hB_{\mu\nu}\text{Tr}[W^{\mu\nu}\Sigma\sigma^3\Sigma^\dagger]~.
\ee
One can easily check that these operators are
independent, hence they are in one-to-one correspondence with
$O_{Z\gamma}$, $O_{ZZ}$ and the $hF_{\mu\nu}F^{\mu\nu}$ interaction
(omitted from the Lagrangian \eqref{eq:lag}) in the unitary gauge. The dimension-3 operator
$h Z^\mu Z_\mu$ can be obtained from $h\text{Tr}[D^\mu\Sigma^\dagger
D_\mu\Sigma]$.  In the linear realization the operator $O_h$ produces
the dimension-3 operator but only in combination with a $Z$-mass term,
so that the $hZZ$ coupling is still given in terms of the physical $Z$
mass as in the SM.  It is because in the linear realization this
dimension-3 operator is effectively absent that we omitted it from the effective
Lagrangian~\eqref{eq:lag}. 

Likewise, one can obtain the $O_{ZJ}=hZ_{\mu}J^\mu$ interactions from
\eqref{eq:lag} in the nonlinear realization of  electroweak
symmetry. For this, one can form the following doublets from the SM
fermions, \beq
\label{}
q_L=\left(\begin{array}{c} u_L\\ d_L\\ \end{array} \right),
~~q_R=\left(\begin{array}{c} u_R\\ d_R\\ \end{array} \right), ~~
\ell_L=\left(\begin{array}{c} \nu_L\\ e_L\\ \end{array} \right), ~~
\ell_R=\left(\begin{array}{c} 0\\ e_R\\ \end{array} \right)~.  
\eeq
Note, that under $U(1)_Y$, the right handed doublets transform
as 
$$q_R\to e^{i\alpha_Y/6} e^{i\alpha_Y\sigma^3/2} q_R~, \qquad
\ell_R\to e^{i\alpha_Y/2} e^{i\alpha_Y\sigma^3/2} \ell_R~.$$ 
Using the latter transformation property, one can straightforwardly
check that the following (independent) operators
\ba
\label{eq:nl2}
ih\bar{ \psi}_L\Sigma \gamma^\mu (D_{\mu}\Sigma)^\dagger\psi_L&+&h.c.~, \quad ih\bar{ \psi}_L\Sigma\sigma^3 \gamma^\mu (D_{\mu}\Sigma)^\dagger\psi_L+h.c.~,\nn\\
ih\bar{ \psi}_R\Sigma^\dagger \gamma^\mu D_{\mu}\Sigma\psi_R&+&h.c.~,
\quad ih\bar{ \psi}_R\sigma_3\Sigma^\dagger \gamma^\mu
D_{\mu}\Sigma\psi_R+h.c.~, 
\ea 
where $\psi$ stands for either quark or lepton fields, are invariant
under $SU(2)_W\times U(1)_Y$. Moreover, in the unitary gauge they lead
precisely to the current interactions, given in~\eqref{eq:lag}, with
the coefficients $c_L$ and $c_R$ independent of each other.

Since the
operators in \eqref{eq:nl1}-\eqref{eq:nl2} only contribute to
interactions that involve the Higgs field, their coefficients are not
constrained by EWPD. Moreover,
the only appreciable constraint on these coefficients comes from the
current bounds on the $h\to \gamma\gamma$ rate, which constrains only one
combination of the operators in \eqref{eq:nl1}. The Lagrangian
\eqref{eq:lag} on the other hand arises from the remaining two
combinations in \eqref{eq:nl1}, as well as the operators in
\eqref{eq:nl2}. These operators are unconstrained by current data.

\section{Three-body Decays}

\subsection{Kinematics and Decay Rates}
In the present subsection we collect analytic expressions for the spin-averaged squared amplitudes $\overline{|\mathcal{M}|^2}$, for various NP contributions to the $h\to Z \ell \bar \ell$ decay.   Consider the three-body decay $h\to Z(p_1) \ell(p_2) \bar
\ell(p_3)$ illustrated in Fig.~\ref{fig:diagram6finalstate}. Following~\cite{Beringer:1900zz}, we define $p_{ij} \equiv p_i + p_j$ and
$m_{ij}^2 \equiv p_{ij}^2$, so that $m_{12}^2 + m_{13}^2 + m_{23}^2 =
M_h^2 + M_Z^2$.  Since the recently discovered resonance has spin-0, the phase space integrals over
angular variables are trivial and the maximum amount of information is
contained in the doubly-differential decay rate,
\begin{equation}
\frac{d^2\Gamma}{dm_{12}^2 dm_{23}^2} = \frac{\overline{|\mathcal{M}|^2}}{256 \pi^3 M_h^3}.
\end{equation}
For a given value of $m_{23}$, the limits on $m^2_{12}$ become 
\be
(m_{12}^2)_{min,~max}=\frac{M_h^2 + M_Z^2 - m_{23}^2 \mp \sqrt{m_{23}^4 - 2 m_{23}^2(M_h^2 + M_Z^2) + (M_h^2 - M_Z^2)^2}}{2},
\ee
with $M_h - M_Z \geq m_{23} \geq 0$.
Similarly, keeping $m_{12}$ fixed, $m_{23}$ can range from $0$ to $(M_H^2 - m_{12}^2)(m_{12}^2 - M_Z)/m_{12}^2$ with $M_h \geq m_{12} \geq M_Z$. 

In order to identify the relative importance of the NP operators from \eqref{eq:lag} for the decays at hand, we will treat them one at a time. For the case of the SM supplemented by a NP operator $O$, the full squared amplitude $\overline{|\mathcal{M}|^2}$ is defined as
\beq
\overline{|\mathcal{M}|^2}=\overline{|\mathcal{M}|^2}_{SM} +\overline{|\mathcal{M}|^2}_{SM-O}+\overline{|\mathcal{M}|^2}_O~,
\eeq
with the last two terms representing the contributions from the interference with the standard model and pure new physics respectively. 
At the leading order in the SM, three-body decays $h\to Z\ell\bar\ell$ proceed through the tree-level $h\to ZZ^*\to Z\ell\bar\ell$ diagram. The spin-averaged squared amplitude for this process is given by the following expression
\begin{equation} \label{eq:sm}
\overline{|\mathcal{M}|^2}_{SM} = \frac{e^4 (g_L^2 + g_R^2)}{s_W^4 c_W^4}~\frac{M_Z^2(2 m_{23}^2 - M_h^2) + m_{12}^2(M_h^2 + M_Z^2 - m_{23}^2) - m_{12}^4}{(m_{23}^2 - M_Z^2)^2},
\end{equation}
where $g_L = -1/2 + s_W^2$ and $g_R = s_W^2$ are the SM couplings of charged leptons to the $Z$ boson. 

The contributions of the operators given in \eqref{eq:lag} to the
$h\to Z\ell\bar\ell$ decays with the final state leptons having
electric charge $Q$, are given as follows (again, we include the
dipole term only for completeness)
%\footnote{The dipole terms in \eqref{eq:lag} can not arise from the flavor $U(3)^5$ - invariant operators. Nevertheless, we include their contributions for completeness. Moreover, the effects of these operators for the processes we are considering should be expected to be rather weak, since they do not interfere with the leading SM diagram.}
\begin{enumerate}
\item{$hZ_{\mu\nu} F^{\mu\nu}$}
\be
\overline{|\mathcal{M}|^2}_{SM-Z\gamma}=\frac{ c_{Z\gamma} Q 8\pi \alpha^3 (g_L+g_R)}{ s^4_W c^4_W}~\frac{m_{23}^2+M_Z^2-M_h^2}{m_{23}^2-M_Z^2}~,\nn
\ee
\be
\overline{|\mathcal{M}|^2}_{Z\gamma}=\frac{ c^2_{Z\gamma} Q^2 \alpha^4}{s^4_W c^4_W}~\frac{2 m_{12}^4+ 2 m^2_{12}( m^2_{23}-M_Z^2-M_h^2) +(m_{23}^2-M_h^2)^2 +2m_{23}^2 M_Z^2+M_Z^4        }{m^2_{23} M_Z^2}~.\nn
\ee
\item{$hZ_\mu J^\mu$}
\be
|\mathcal{M}|^2_{SM-ZJ}= \frac{e^4(c_L g_L + c_R g_R)}{s_W^4 c_W^4}~\frac{M_Z^2(2 m_{23}^2 - M_h^2) + m_{12}^2(M_h^2 + M_Z^2 - m_{23}^2) - m_{12}^4}{(m_{23}^2 - M_Z^2) M_Z^2},\nn
\ee
\be
|\mathcal{M}|^2_{ZJ}= \frac{e^4(c_L^2 + c_R^2)}{4 s_W^4 c_W^4}~\frac{M_Z^2(2 m_{23}^2 - M_h^2) + m_{12}^2(M_h^2 + M_Z^2 - m_{23}^2) - m_{12}^4}{M_Z^4}~.  \nn
\ee
\item{$hZ_{\mu\nu} Z^{\mu\nu}$}
\be
\overline{|\mathcal{M}|^2}_{SM-ZZ}=\frac{ c_{ZZ} 16\pi\alpha^3 (g^2_L+g^2_R)}{ s^6_W c^6_W}~\frac{m^2_{23}(m_{23}^2+M_Z^2-M_h^2)}{(m_{23}^2-M_Z^2)^2}~,\nn
\ee
\ba
\overline{|\mathcal{M}|^2}_{ZZ}=\frac{ c^2_{ZZ} 4\alpha^4(g^2_L+g^2_R)}{ s^8_W c^8_W}~\bigg(m^2_{23} (2m^4_{12}+2m^2_{12}(m^2_{23}-M_Z^2-M_h^2) \nn \\+(m_{23}^2-M_h^2)^2 +2m_{23}^2 M_Z^2+M_Z^4  \bigg) / M_Z^2(m^2_{23}-M_Z^2)^2~.\nn
\ea
\item{$h\bar\ell\sigma^{\mu\nu}\ell Z_{\mu\nu}$}
\be
|\mathcal{M}|^2_{SM-\text{Dipole}}=0~,\nn
\ee
\be
|\mathcal{M}|^2_{Dipole}= \frac{c^2_{\ell Z}e^6}{2 s_W^6 c_W^6}~\frac{-2m^2_{12}+2m^2_{12}(M_h^2-m^2_{23}-M_Z^2)+M_Z^2(m_{23}^2-2 M_h^2)}{M_Z^4}~.  \nn
\ee
\end{enumerate}

Integrating the SM doubly-differential rate with respect to $m_{12}^2$, yields
\begin{equation}
\frac{d\Gamma_{SM}}{dm_{23}^2} = \frac{e^4  (g_L^2 + g_R^2) 2 m_{23} |\mathbf{p_1}|}{1536 \pi ^3 M_h^3 s_W^4 c_W^4} ~\frac{m_{23}^4 - 2 m_{23}^2(M_h^2 - 5 M_Z^2) + (M_h^2 - M_Z^2)^2}{(m_{23}^2 - M_Z^2)^2},
\end{equation}
where $\mathbf{p_1}$ is the momentum of the $Z$ in the $\ell\bar{\ell}$ center-of-mass frame, $2 m_{23} |\mathbf{p_1}| = \sqrt{m_{23}^4 - 2 m_{23}^2(M_h^2 + M_Z^2) + (M_h^2 - M_Z^2)^2}$.  
The corresponding quantities in the presence of the operators $O_{\gamma Z}$ and $O_{ZJ}$, including interference with the SM, are respectively given as follows
\begin{align} \label{eq:ZAdiff}
\frac{d\Gamma_{Z\gamma}}{dm_{23}^2} &= \frac{c_{Z\gamma} Q \alpha^3 2 m_{23} |\mathbf{p_1}|}{32 \pi^2 M_h^3 s_W^4 c_W^4}\left[(g_L + g_R)\left(\frac{m_{23}^2 - M_h^2 + M_Z^2}{m_{23}^2 - M_Z^2}\right)\right. \\
&\left.+ \frac{c_{Z\gamma} Q \alpha}{12 \pi}\left(\frac{m_{23}^4 - 2m_{23}^2(M_h^2 - 2M_Z^2) + (M_h^2 - M_Z^2)^2}{m_{23}^2 M_Z^2}\right)\right] ,\nn\\
\frac{d\Gamma_{ZJ}}{dm_{23}^2} &= \frac{d\Gamma_{SM}}{dm_{23}^2} \left(\frac{c_L g_L + c_R g_R}{g_L^2 + g_R^2}~\frac{m_{23}^2 - M_Z^2}{M_Z^2} + \frac{1}{4}\frac{c_L^2 + c_R^2}{g_L^2 + g_R^2}~\frac{(m_{23}^2 - M_Z^2)^2}{M_Z^4} \right).\nn
\end{align} 
The complete differential decay rate is obtained by summing over the individual pieces (as well as including interference between the various NP contributions),
\begin{equation}
\frac{d\Gamma_{h \rightarrow Z \ell \bar{\ell}}}{dm_{23}^2} = \frac{d\Gamma_{SM}}{dm_{23}^2}  + \frac{d\Gamma_{ZJ}}{dm_{23}^2} + \frac{d\Gamma_{Z\gamma}}{dm_{23}^2} + \cdots.
\end{equation}

\subsection{Discussion}
\label{sec:dis}
In this subsection, we have identified the
regions in the three-body phase space where the effects of new physics can become
especially pronounced and could be detectable by future experiments.  We define signal strength in the standard way,
\begin{equation}
\mu = \frac{\sigma_{SM+NP} \times \text{Br}_{SM+NP}}{\sigma_{SM} \times \text{Br}_{SM}}.  
\end{equation}
None of the models we consider modify the Higgs-like production rate significantly.  In addition, $\mathcal{O}_{ZJ}$ does not affect the total width of the new resonance for the parameters chosen, while modifications of the total width due to $\mathcal{O}_{Z\gamma}$ are loop suppressed.  Therefore, to a good approximation, the signal strength is given by $\mu = \Gamma_{SM+NP} / \Gamma_{SM}$.  We generalize this concept to that of the \textit{differential} signal strength with the following definition, $$\mu(m_{23}^2) = (d\Gamma_{SM+NP}/dm_{23}^2) / (d\Gamma_{SM}/dm_{23}^2).$$ 

\begin{figure}
  \centering
  \subfloat{\label{fig:rate}\includegraphics[width=0.5\textwidth]{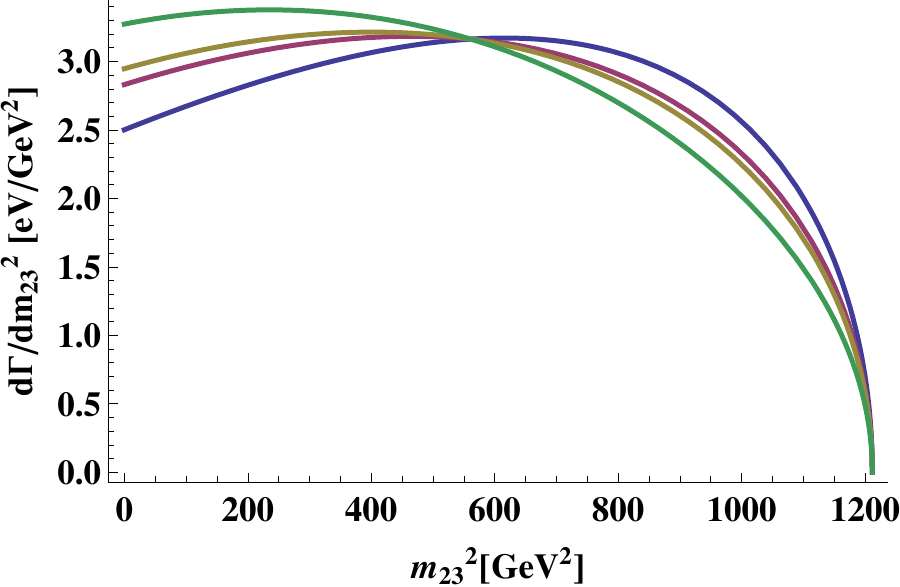}}
  \subfloat{\label{fig:signal}\includegraphics[width=0.5\textwidth]{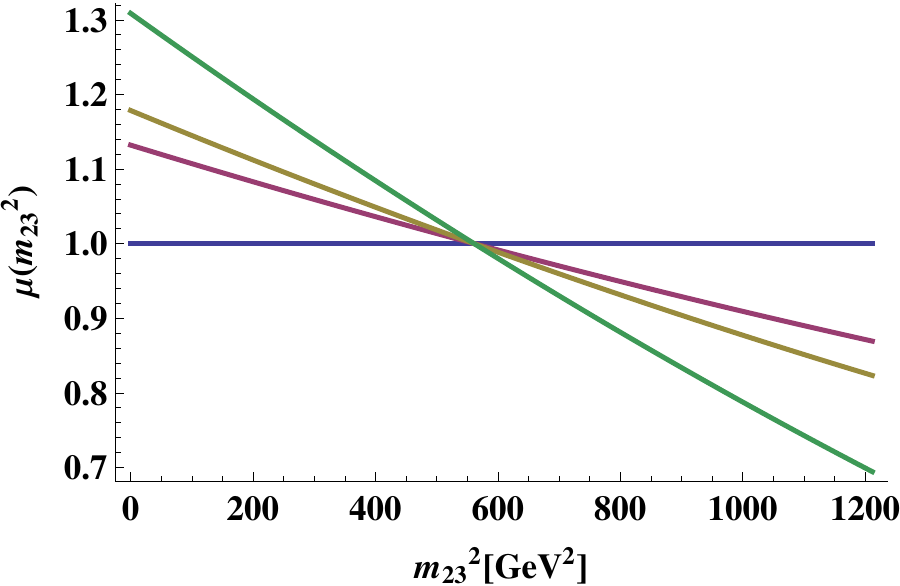}}
  \caption{Contributions to $h \to Z \ell \bar{\ell}$ from
    $\mathcal{O}_{ZJ}$.  The differential decay rate and differential
    signal strength as a function of $m_{23}^2$ are shown on the left
    and right respectively.  The curves correspond to the SM (blue);
    $c_R = 0.99,\, c_L = 0$ (red); $c_L = -1.15,\, c_R = 0$ (yellow);
    and $c_R = -c_L = 1.07$ (green). $\mu = 1$ in each of these
    cases.}
  \label{fig:current}
\end{figure}

Fig.~\ref{fig:current} shows differential decay rate (left) and differential signal strength (right) as a function of $m_{23}^2$ for the current interaction $O_{ZJ}$.  The values of the coefficients in Fig.~\ref{fig:current} are all within the $1\sigma$ bounds from EWPD in Table~\ref{HSas}.  All of these curves do not add or subtract from the total SM decay rate, $\mu = 1$, and yet can differ from the leading order (LO) SM prediction by as much as 30\% in the differential decay rate.  We do not expect next-to-leading order (NLO) SM corrections to significantly affect this result for the values of $m_{23}$ above a few GeV, see Fig.~\ref{fig:smza}. Allowing the total signal strength to be as large as the 1$\sigma$ bound on $h \to Z Z^{\star}$ in~\cite{Giardino:2013bma}, $\mu = 1.00 \pm 0.13$, the effect can become as large as 50\%. 
Searching for this effect in $h \rightarrow Z q \bar{q}$ ($q = b,c$)
at a linear collider may provide an even better window to observe the
effects of NP.  The reason being that the bounds from $Z$-pole
experiments are weaker on heavy quarks than they are on leptons, and
at least currently, the bounds from Higgs data are weaker on heavy
quarks than they are EW gauge bosons.
%\begin{figure}
%  \centering
%\includegraphics[width=0.5\textwidth]{curLEP.pdf}
%   \caption{Contributions from $\bar{\mathcal{O}}_{hf}$ to the differential signal strength as a function of $m_{23}^2$.  SM (blue), $\bar{\mathcal{O}}_{h\mu}$ with $C_R = - C_L = 0.003$ (red), $\bar{\mathcal{O}}_{hb}$ with $C_R = -0.7,\, C_L = 0$ (yellow), $\bar{\mathcal{O}}_{hc}$ with $C_R = C_L = -0.02$ (green)}
% \label{fig:Ocurrent}
%   \end{figure}

In order to fully demonstrate that three-body decays of the higgs bosons are good probe of physics beyond the SM, a semi-realistic detector simulation should be performed. We consider that to be beyond the scope of this proof of principle work. However, as a first step towards that goal we use FeynRules~\cite{Christensen:2008py} to implement $O_{ZJ}$ in MadGraph 5~\cite{Alwall:2011uj} to simulate $pp \to h \to 4\ell$ at LHC8. $1.1\cdot 10^5$ events are generated for a given benchmark point from Fig.~\ref{fig:current} with the renormalization and factorization scales set to $m_h$. MSTW2008 LO PDFs~\cite{Martin:2009iq} are used, and no cuts are placed on the final state leptons. 

\begin{figure}
  \centering
  \subfloat{\label{fig:ratemg}\includegraphics[width=0.5\textwidth]{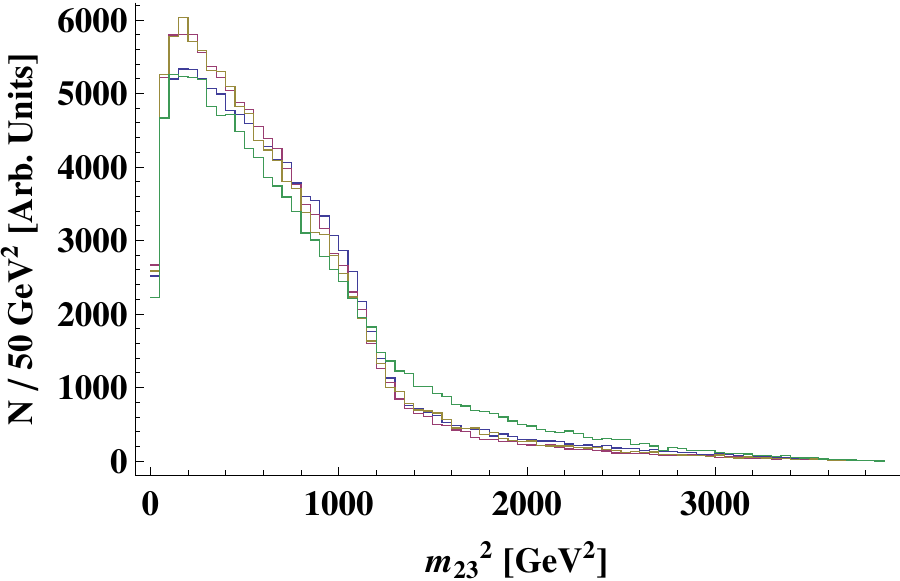}}
  \subfloat{\label{fig:signalmg}\includegraphics[width=0.5\textwidth]{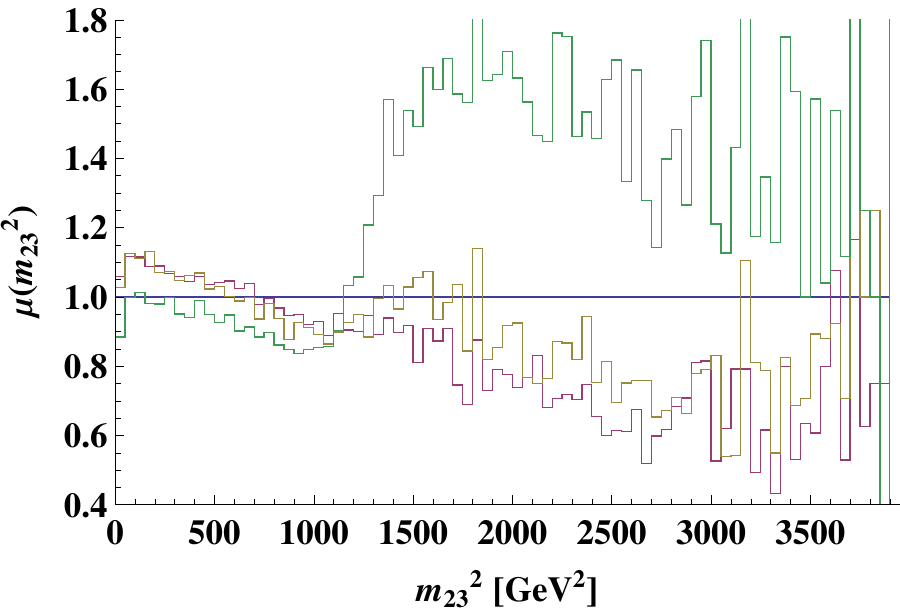}}
  \caption{Contributions to $pp \to h \to 4\ell$ at LHC8 from
    $O_{ZJ}$.  The differential production rate and differential
    signal strength as a function of $m_{23}^2$ are shown on the left
    and right respectively.  The curves correspond to the SM (blue);
    $c_R = 0.92,\, c_L = 0$ (red); $c_L = -1.09,\, c_R = 0$ (yellow);
    and $c_R = -c_L = 0.92$ (green). $\mu = 1$ within statistical uncertainty in each of these
    cases.}
  \label{fig:currentmg}
\end{figure}

\begin{figure}
  \centering
  \subfloat{\label{fig:smin}\includegraphics[width=0.5\textwidth]{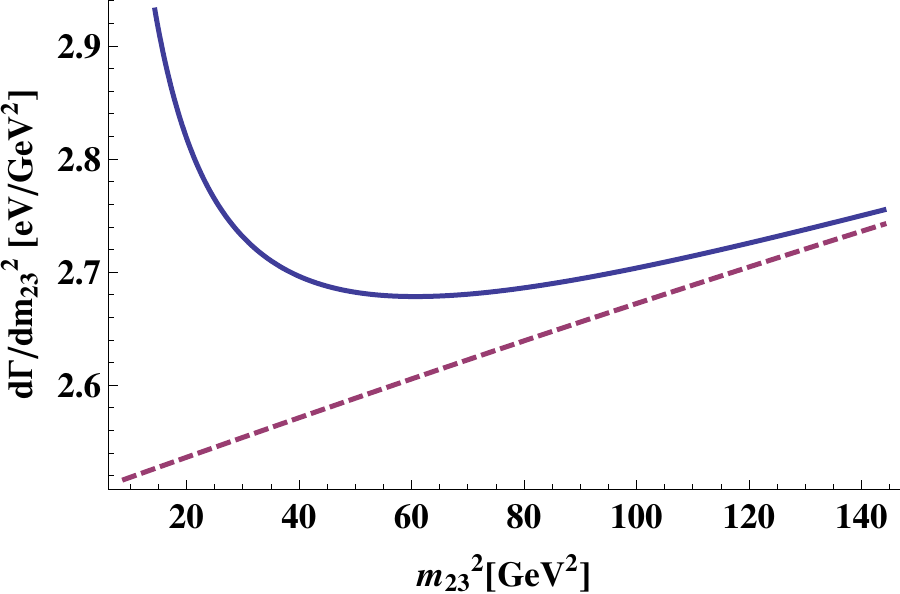}}
  \subfloat{\label{fig:smout}\includegraphics[width=0.5\textwidth]{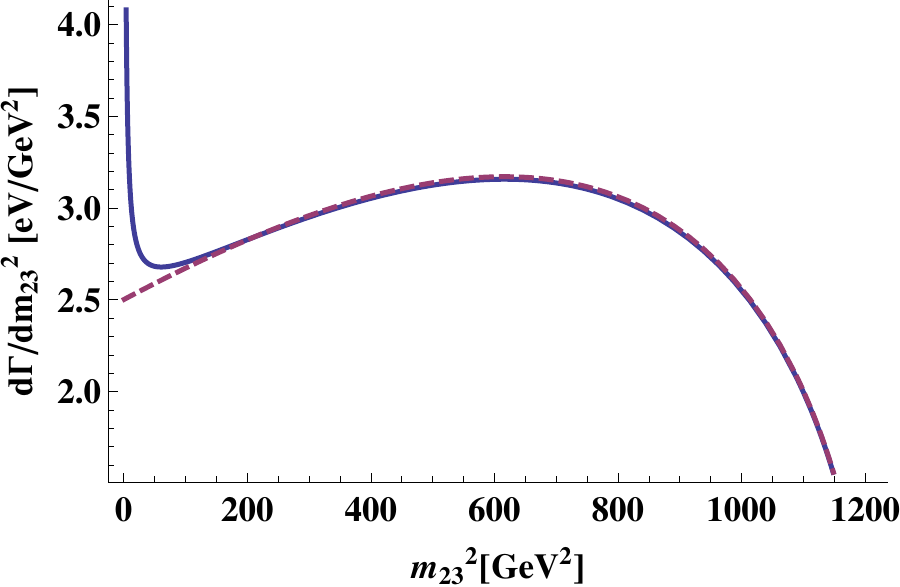}}
  \caption{Differential rates for $h \rightarrow Z \ell \bar{\ell}$ in
    the SM.  Shown is red is the LO process, which has a tree-level $Z
    Z^{\star}$ intermediate state. The blue curve includes the
    dominant NLO corrections, which come from the loop-induced $Z
    \gamma^{\star}$ intermediate state.}
  \label{fig:smza}
\end{figure}

The results of our simulation are shown in
Fig.~\ref{fig:currentmg}. The differential production rate and
differential signal strength as a function of $m_{23}^2$ are shown on
the left and right respectively.  The curves correspond to the SM
(blue); $c_R = 0.92,\, c_L = 0$ (red); $c_L = -1.09,\, c_R = 0$~(yellow); and $c_R = -c_L = 0.92$ (green). $\mu = 1$ within
statistical uncertainty in each of these cases.
 
There are effects captured in the MadGraph simulation that are not
present in the analytic calculation. Most noticeably, the distribution
of events with respect to $m_{23}$ does not end at $m_h - m_Z$, but
instead at $m_h$ since there is a tail due to the both $Z$'s being
off-shell. The fractional deviation from the SM in the tail of the
distribution is significantly larger than in the body of the
distribution, but the rate is smaller.  In addition, there are
interference effects in the four-body final state that are not present
in the three-body final state. These effects shift the values of the
couplings needed to keep the signal strength for $h \to 4\ell$ at the
SM value, $\mu(h \to ZZ) = 1$. The magnitude of the couplings required
in this analysis are smaller than those used in
Figure~\ref{fig:current}.

A particularly interesting operator in the context of three-body Higgs
decays is $O_{Z\gamma}$.  In the massless fermion limit, the decay
rate for $h \rightarrow Z \gamma^{\star} \rightarrow Z \ell
\bar{\ell}$ diverges logarithmically as $m_{23} \rightarrow 0$, due to
the integration over phase space.\footnote{Upon inclusion of the
  final-state fermion masses, the divergence is removed due to the
  lower bound on the kinematic variable $(m_{23}^2)_{\text{min}} = 4
  m_f^2$.} This formal divergence cancels against the IR divergence
arising due to the interference between tree level diagram for the
decay $h \rightarrow Z \gamma$ and its one-loop photon self-energy
correction. The two processes are indistinguishable 
for $m_{23}$ below detector resolution. For low  values of $m_{23}$
 this results
in a large enhancement of the effects of the given operator. An
experiment with enough resolution, perhaps around a few GeV, can therefore
become extremely sensitive to the differential decay rate. The
enhancement can be clearly seen in Fig. \ref{fig:smza}, where we plot
the approximate NLO SM (blue), and LO SM (red) differential rates, in
the presence of the SM loop-induced $h\rightarrow Z \gamma^{\star}
\rightarrow Z \ell \bar{\ell}$ decay in addition to the tree-level SM
diagram.
% Experimentally what happens is below a threshold energy, one photon is not distinguishable from a pair of collinear leptons.  Because the divergence is canceled exactly at $m_{23} = 0$ ($d\Gamma_{h \rightarrow Z \ell \bar{\ell}} / dm_{23}^2 \supset \Gamma_{h \rightarrow Z \gamma} \delta(m_{23}^2 = 0)$) 
The current analyses~\cite{CMS-PAS-HIG-13-002, ATLAS-CONF-2013-013} of
$h \rightarrow 4\ell$ only consider $(m_{23})_{\text{min}}$ above
12~GeV, and are less sensitive to this effect. A dedicated experiment,
either at the LHC or at a future precision facility, with lower
$m_{23}$ threshold could easily capture this physics.

%Alternatively, it might not be possible to observe this effect at the LHC, perhaps the backgrounds are too high in this low mass region of phase space.  
%In either case, we are hopeful that a precision machine, such as a linear collider will be able to capture this effect and provide a clean environment for its studies.
%\begin{figure}
%  \centering
% \subfloat{\label{fig:rateKM}\includegraphics[width=0.5\textwidth]{ZFmu.pdf}} 
%  \subfloat{\label{fig:signalKM}\includegraphics[width=0.5\textwidth]{ZZmu.pdf}}
%   \caption{Differential signal strength for $\mathcal{O}_{Z\gamma}$ (left) and $\mathcal{O}_{ZZ}$ (right) as a function of $m_{23}^2$.  $C_{Z\gamma}$ and $C_{ZZ}$ vary from 1.0 to 2.0 in steps of 0.2.  The SM corresponds to $\mu = 1$.}
% \label{fig:kinmix}
%   \end{figure}

\begin{figure}
  \centering
  \subfloat{\label{fig:rin}\includegraphics[width=0.5\textwidth]{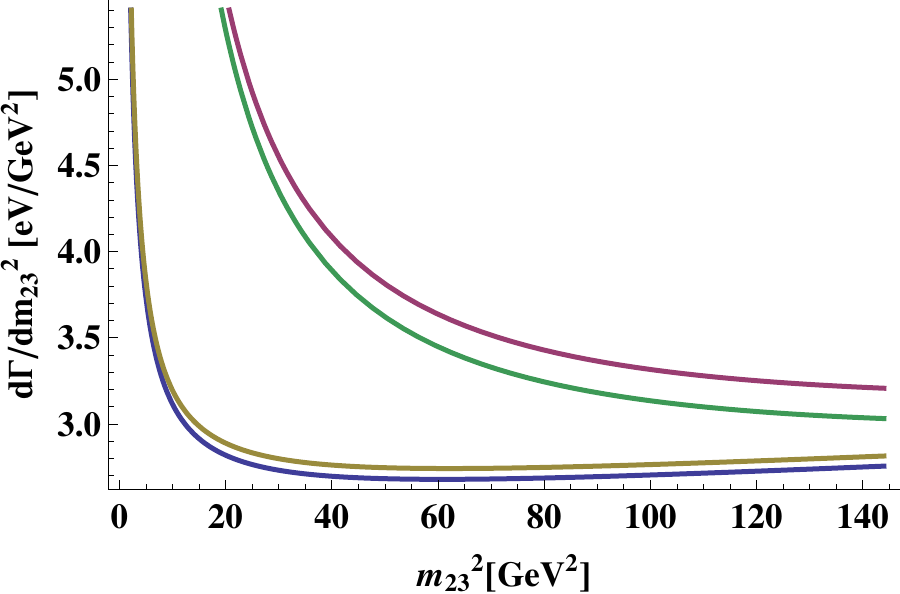}}
  \subfloat{\label{fig:rout}\includegraphics[width=0.5\textwidth]{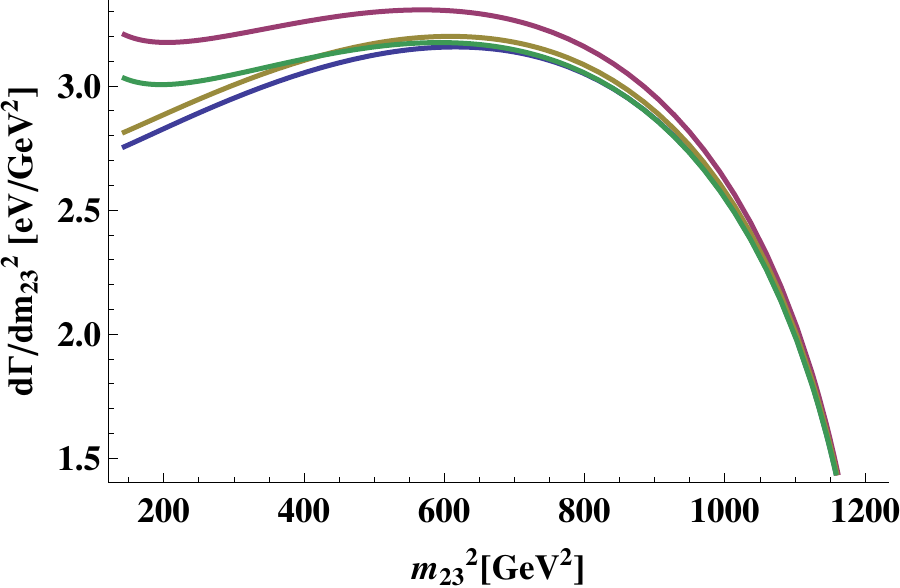}} \\
  \subfloat{\label{fig:min}\includegraphics[width=0.5\textwidth]{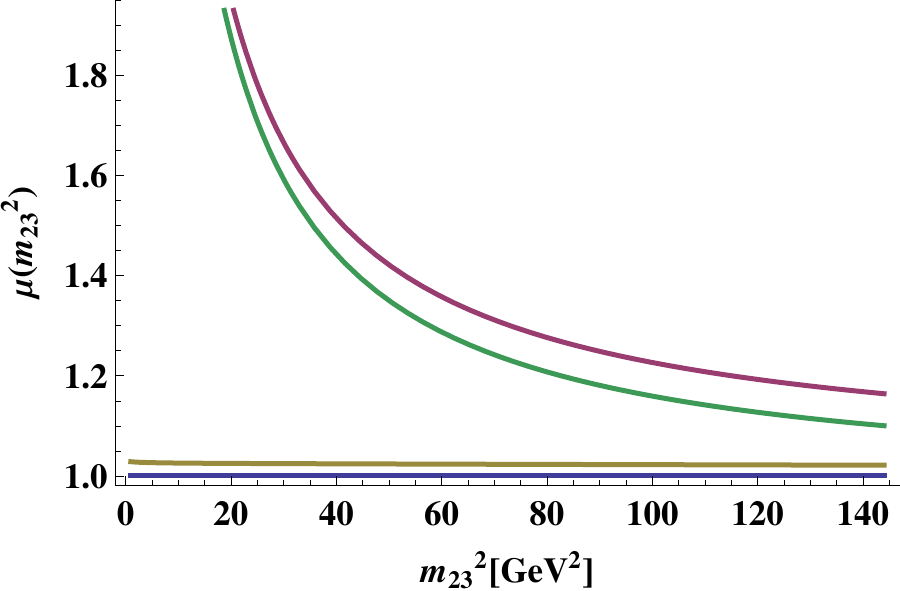}}
  \subfloat{\label{fig:mout}\includegraphics[width=0.5\textwidth]{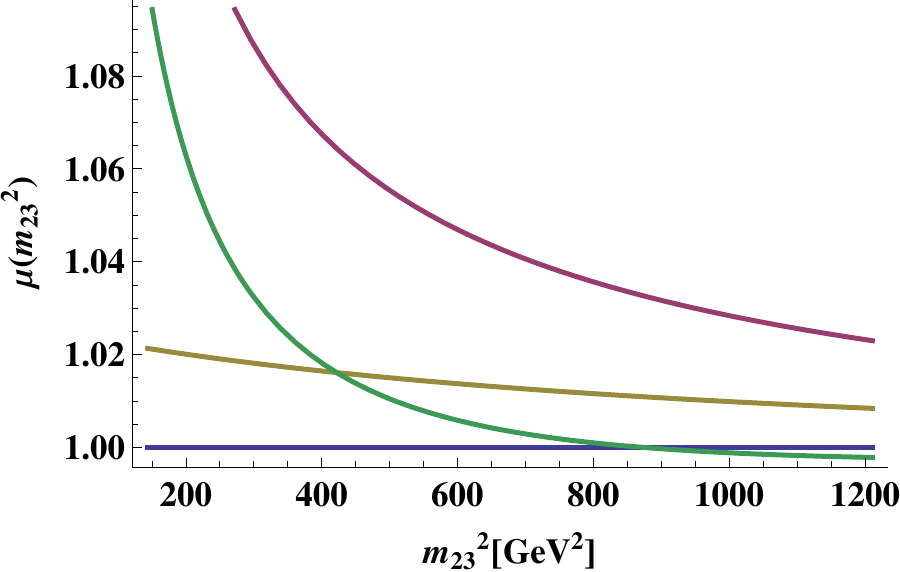}}
  \caption{The differential rates and signal strengths in the SM,
    supplemented by a NP contribution to the operator
    $\mathcal{O}_{Z\gamma}$. The curves correspond to: SM including
    $c_{Z\gamma,\,SM}$ (blue); $c_{Z\gamma,\,NP} = 20$ (red);
    $c_{Z\gamma,\,NP} = 10$ (yellow); $c_{Z\gamma,\,NP} = -10$
    (green). In order to emphasize the enhancement of NP effects for
    the lower values of $m_{23}$, we restrict $m_{23}$ to be less than
    12 GeV in the plots on the left, while $m_{23} > 12$ GeV is
    considered for the plots on the right.}
 \label{fig:smza1}
\end{figure}

As emphasized above, the process $h \to Z \gamma^{\star} \to Z f
\bar{f}$ occurs in the SM as well as in NP scenarios.  The SM
contributes to $O_{Z\gamma}$ with the coefficient $c_{Z\gamma,\,SM}
\approx -4.82$. In addition, the SM loop also contributes to
dimension-8 operators, such as $v h Z_{\mu\nu} \partial^2 F^{\mu\nu}$.
These terms are finite as $m_{23} \to 0$, or to put it differently,
these terms are suppressed by powers of $m_{23}^2 / M_Z^2$, so we may
neglect them.  We also neglect loop-corrections to $h \to Z Z^{\star}
\to Z f \bar{f}$.  Even though these corrections are formally the same
order as $h \to Z \gamma^{\star} \to Z f \bar{f}$, they are
sub-leading in the kinematic region of interest, {\it i.e.}, low values of
$m_{23}$.

Inspecting Fig.~4 of Ref.~\cite{CMS-PAS-HIG-13-006}, the current
bound on $h \to Z \gamma$ is roughly 9 times the SM rate at 95\% CL,
which translates into $-10 \lsim c_{Z\gamma,\,NP} \lsim 20$.  Using
Eq.~\eqref{eq:tbase} and the $1\sigma$ bounds in Table~\ref{HSas},
yields $-267 \le c_{Z\gamma,\,NP} \le 64$. Taking only these weak
constraints into account, the effect on the differential rate for $h
\to Z \ell \bar{\ell}$ can be rather dramatic, as can be seen in
Fig.~\ref{fig:smza}.

It is natural to wonder if the values chosen for $c_{Z\gamma}$ modify
other observables since they seem rather large.  Even for the largest
value of the coefficient, $c_{Z\gamma}=20$, the integrated signal
strength for $h \to Z \ell \bar{\ell}$ is quite close to the SM
prediction: $\mu = 1.06$ with the cut $m_{23}>12~$GeV (which is what
the current analyses
employ~\cite{CMS-PAS-HIG-13-002, ATLAS-CONF-2013-013}), and becomes $\mu = 1.13$ if the cut is taken to be $m_{23}>1~
$GeV. Both of these values are within the 1$\sigma$ bound on $h \to Z
Z^{\star}$ from Ref.~\cite{Giardino:2013bma}.

Furthermore, in UV completions of the effective theory the coefficient
$c_{Z\gamma}$ is a combination of coefficients in the Han and Skiba
basis. For example, in Manohar's model~\cite{Manohar:2013rga} it is
given in terms of $a_{WW}$ and $a_{BB}$. Therefore it is reasonable to
expect the UV completion will generate $c_{ZZ}$ and $c_{WW}$ of
comparable magnitude to $c_{Z\gamma}$.\footnote{$c_{WW}$ is defined
  through $$\Delta\mathcal{L}_{NP}=\frac{\alpha}{4
    \pi}\left(\frac{2c_{WW}}{s_W^2} W^+_{\mu\nu} W^{-\mu\nu} \right)\frac{h}{v}$$} 
For $c_{ZZ} = \{-10,10,20\}$ we
find the signal strength for $h \to Z \ell \bar{\ell}$ is $\{1.03,
0.97, 0.94\}$. Similarly, for $c_{WW} = \{-10,10,20\}$ we find the
signal strength for $h \to W f \bar{f}^{\prime}$ is $\{1.05, 0.95,
0.91\}$. These signal strengths were computed using only the tree
level piece of the SM contribution, and it is likely that the NLO SM
contributions are of the same order as the NP contributions. For
example, in the SM the effective coefficients (that reproduce the SM
amplitude as if it resulted from the EFT operators) are
$c_{\gamma\gamma} = -6.98$ and $c_{Z\gamma} =
-4.82$~\cite{Djouadi:2005gi}, indeed comparable to the NP
contributions.

\section{Conclusions}
The $125$ GeV resonance, observed at the LHC closely resembles the
Higgs of the minimal standard model. If the production rates and the
partial widths of the resonance continue to converge towards their SM
values, and no direct evidence for additional degrees of freedom is
found within the LHC range, the identification of indirect probes of
new physics will become increasingly important.

In the present work we have concentrated on one possible such probe,
the three-body decays of the Higgs boson, $h\to Z\ell\bar\ell$. We
have shown that these processes represent a promising avenue towards
capturing the effects of new physics. Moreover, we have identified the
regions in the three-body phase space in which these effects become
especially pronounced and could be detected at precision facilities,
or even via dedicated experiments at the LHC.

Many interesting questions remain. In particular, it would be
important to understand implications of concrete NP models in the
context of the above analysis, as well as study ways in which form
factors of the Higgs can be probed via various interactions in the
standard model or its extensions. These questions represent
the subject of our ongoing investigation \cite{paper}.

\textbf{Note Added:} As this work was being
   completed, Ref.~\cite{Isidori:2013cla} appeared  with some overlap
   with our work. Our work in progress was mentioned by one of us at a
   presentation at the Higgs Factory Muon Collider Workshop, UCLA on
   March 21, 2013; see pp.17--21 in
   \url{https://hepconf.physics.ucla.edu/higgs2013/talks/grinstein.pdf}. 

\begin{acknowledgments}
We thank W. Skiba for helpful discussions regarding the analysis in
\cite{Han:2004az}, and P. Uttayarat useful comments. This work has been supported in part by the U.S. Department of Energy
under contract No. DE-SC0009919. %DOE-FG03-97ER40546
\end{acknowledgments}

\bibliography{Higgs3body}

%
%
%%%%%%%%%%%%%%%%%%%%%%%%%%%%%%%%%%%%%%%%%%%%%
%%%%%%%%%%%%%%%%%%%%%%%%%%%%%%%%%%%%%%%%%%%%%
%%%%%%%%%%%%%%%%%%%%%%%%%%%%%%%%%%%%%%%%%%%%%%5
\end{document}